\documentclass[12pt]{article}
\usepackage{epsfig}

\def\beq{\begin{equation}}
\def\eeq#1{\label{#1}\end{equation}}
\def\eeqn{\end{equation}}

\def\beqa{\begin{eqnarray}}
\def\eeqa#1{\label{#1}\end{eqnarray}}
\def\eeqan{\end{eqnarray}}

\let\bar=\overbar

\def\Dslash{\not{\hbox{\kern-4pt $D$}}}
\def\dslash{\not{\hbox{\kern-2pt $\del$}}}

\def\msb{{\bar{\ssstyle M \kern -1pt S}}}

\def\Title#1{\begin{center} {\Large {\bf #1} } \end{center}}

\begin{document}

\def\mbc{M_{bc}}
\def\de{\Delta E}
\def\bsdsds{B_s^0\to D_s^{(*)-}D_s^{(*)+}}
\def\dsstdsst{D_s^{*-}D_s^{*+}}
\def\dsstds{D_s^{*\pm}D_s^{*\mp}}
\def\dsds{D_s^{-}D_s^{+}}
\def\fl{f^{}_L}
\def\bs{B_s}
\def\bsst{B^{*}_s}
\def\bsbar{\overline{B}{}^{}_s}
\def\bsbarst{\overline{B}{}^{\,*}_s}
\def\bbar{\overline{B}{}^{\,0}}
\def\bbbar{$B^0$-$\bbar$}
\def\dgs{\Delta\Gamma^{}_s}
\def\dgcp{\Delta\Gamma^{CP}_s}
\def\gs{\Gamma^{}_s}

\def\kstz{K^{*0}}
\def\kstp{K^{*+}}

\def\mevm{~MeV/$c^2$\/}
\def\mevp{~MeV/$c$\/}
\def\meve{~MeV}
\def\gevm{~GeV/$c^2$\/}
\def\gevp{~GeV/$c$\/}
\def\geve{~GeV}

\Title{$\Delta\Gamma_s$ Measurement at the $\Upsilon$(5S) from Belle}

\bigskip\bigskip


\begin{raggedright}  

{\it Sevda Esen\\
Department of Physics\\
University of Cincinnati\\
Cincinnati, OH, 45220 USA}
\bigskip\bigskip

Proceedings of CKM 2012, the 7th International Workshop on the CKM Unitarity Triangle, University of Cincinnati, USA, 28 September - 2 October 2012

\end{raggedright}

\section{Introduction}
Using the full Belle $\Upsilon(5S)$ data sample of 121~fb$^{-1}$ 
we have measured exclusive branching fractions for the decays 
$B_s^0 \to D_s^{(*)+}D_s^{(*)-}$ ~\cite{Esen-full}. 
Assuming these decay modes saturate decays to $CP$-even final states, 
the branching fraction determines the relative width difference 
between the $CP$-odd and -even eigenstates of the $B_s$.

\section{Event Selection}
 

At the $\Upsilon(5S)$ resonance, the total number of $\bs\bsbar$ pairs produced is 
$N^{}_{\bs\bsbar} =  (7.11\pm 1.30)\times 10^6$.
Three production modes are kinematically allowed: $\bs\bsbar$, 
$\bs\bsbarst$ (or $\bsst\bsbar$), and $\bsst\bsbarst$, where the $\bsst$ decays via $\bsst\to\bs\gamma$.
In this analysis we obtain the signal yield for the last mode, 
for which the fraction ($f^{}_{B^*_s\overline{B}{}^{\,*}_s}$)
is $0.870\,\pm 0.017$~\cite{Louvot-full}.

We reconstruct $B^0_s\to D^{*+}_s D^{*-}_s$, $D^{*\pm}_s D^{\mp}_s$, 
and $D^{+}_s D^{-}_s$ decays in which $D^{*+}_s\to D^+_s\gamma$ and
$D^+_s\to \phi(\to K^+K^-)\, \pi^+$,
$K^0_S(\to \pi^+\pi^-)\,K^+$,
$\kstz(\to K^+\pi^-)\, K^+$,
$\phi(\to K^+K^-)\, \rho^+(\to \pi^+\pi^0)$,
$K^0_S(\to \pi^+\pi^-)\,\kstp(\to K^0_S\,\pi^+)$, and 
$\kstz(\to K^+\pi^-)\, \kstp(\to K^0_S\,\pi^+)$~\cite{charge-conjugates}.

We select $\bs$ candidates
using two quantities evaluated in the 
CM frame: the beam-energy-constrained mass $\mbc=\sqrt{E^2_{\rm beam} - p^2_B}$ 
and the energy difference $\de= E^{}_B-E^{}_{\rm beam}$, where 
$p^{}_B$ and $E^{}_B$ are the reconstructed momentum and energy 
of the $\bs$ candidate and $E_{\rm beam}$ is the beam energy. 
The fit region is selected to be
$\mbc$ in $[5.25,5.45]$\gevm\ and $\de$ in $[-0.15,0.10]$\geve.
Because the $\gamma$ from $\bsst\to\bs\gamma$ is not reconstructed,
the modes $\Upsilon(5S)\to\bs\bsbar$, $\bs\bsbarst$ and 
$\bsst\bsbarst$ are well-separated in $\mbc$ and $\de$.
We expect only small contributions from $\bs\bsbar$ 
and $\bs\bsbarst$ events and fix these contributions 
relative to $\bsst\bsbarst$ according to our measurement 
using $B_s^0 \to D^-_s \pi^+$ decays~\cite{Louvot-full}. 

Approximately half of selected events contain multiple candidates, 
typically due to random photons incorrectly assigned 
as $D_s^*$ daughter photons. 
For these events we select the candidate that minimizes the quantity
$
\frac{1}{(2+N)}\,\biggl\{
\sum_{D^{}_s} \left[ {(M^{}_{\widetilde{D}_s} - 
M^{}_{D_s})}/{\sigma^{}_M}\right]^2 +
\sum_{D_s^{*}} \left[ 
{\Delta\widetilde{M} - \Delta M)}/
{\sigma^{}_{\Delta M}}\right]^2\biggr\}\,,
$
where 
$\Delta\widetilde{M}=M^{}_{\widetilde{D}_s^+\gamma} - M^{}_{\widetilde{D}_s^+}$ 
and $\Delta M =M^{}_{D^*_s}-M^{}_{D^{}_s}$.
The summations run over the two $D^+_s$ daughters and the 
$N$ ($=\!0,1,2$) $D^{*+}_s$ daughters 
of the $B^0_s$ candidate. 

Background from $e^+e^-\to q\bar{q}~(q=u,d,s,c)$ continuum 
events are rejected using a  Fisher discriminant based on a set of modified Fox-Wolfram 
moments~\cite{KSFW}. 
The remaining background consists of 
$\Upsilon(5S)\to B^{(*)}_s\overline{B}{}^{(*)}_s\to D^+_s X$,
$\Upsilon(5S)\to B\overline{B}X$, and
$B^{}_s\to D^\pm_{sJ}(2317)D^{(*)}_s$,
$B^{}_s\to D^\pm_{sJ}(2460)D^{(*)}_s$, and
$B^{}_s\to D^\pm_{s}D^\mp_s\pi^0$ decays. 
The last three processes
are estimated to be small 
using analogous $B^{}_d\to D^\pm_{sJ}D^{(*)}$ branching frations and considered 
 when evaluating systematic uncertainty due to backgrounds.


We measure signal yields by performing a
two-dimensional extended unbinned maximum-likelihood 
fit to the $\mbc$-$\de$ distributions. For each sample, 
we include probability density functions (PDFs) for signal 
and background. 

The signal PDFs have three components: correctly reconstructed (CR) decays; 
``wrong combination'' (WC) decays in which a non-signal track or $\gamma$ is 
included in place of a true daughter track or $\gamma$; and ``cross-feed'' (CF) 
decays in which a $D_s^{*\mp}$ is not fully reconstructed or fake.  
In the former case, the $\gamma$ from
$D^{*+}_s\to D^+_s\gamma$ is lost and $\de$ is shifted down
by $100\!-\!150$\meve; this is called ``CF-down.'' In the latter
case, an extraneous $\gamma$ is included and $\de$ is shifted 
up by a similar amount; this is called ``CF-up.'' 
All signal shape parameters are taken from MC simulation and 
calibrated using $B^0_s\to D^{(*)-}_s \pi^+$ and $B^0\to D^{(*)+}_s D^-$ decays. 
The fractions of WC and CF-down events within the fit region are taken from MC. 
The fractions of CF-up events are floated as they are difficult 
to simulate accurately (i.e., many $B_s^0$ partial widths are 
unmeasured). As the CF-down fractions are fixed, the separate
$D^+_s D^-_s$, $D^{*\pm}_s D^{\mp}_s$, and $D^{*+}_s D^{*-}_s$
samples must be fitted simultaneously.

\begin{figure}[bth]
\begin{center}
\epsfig{file=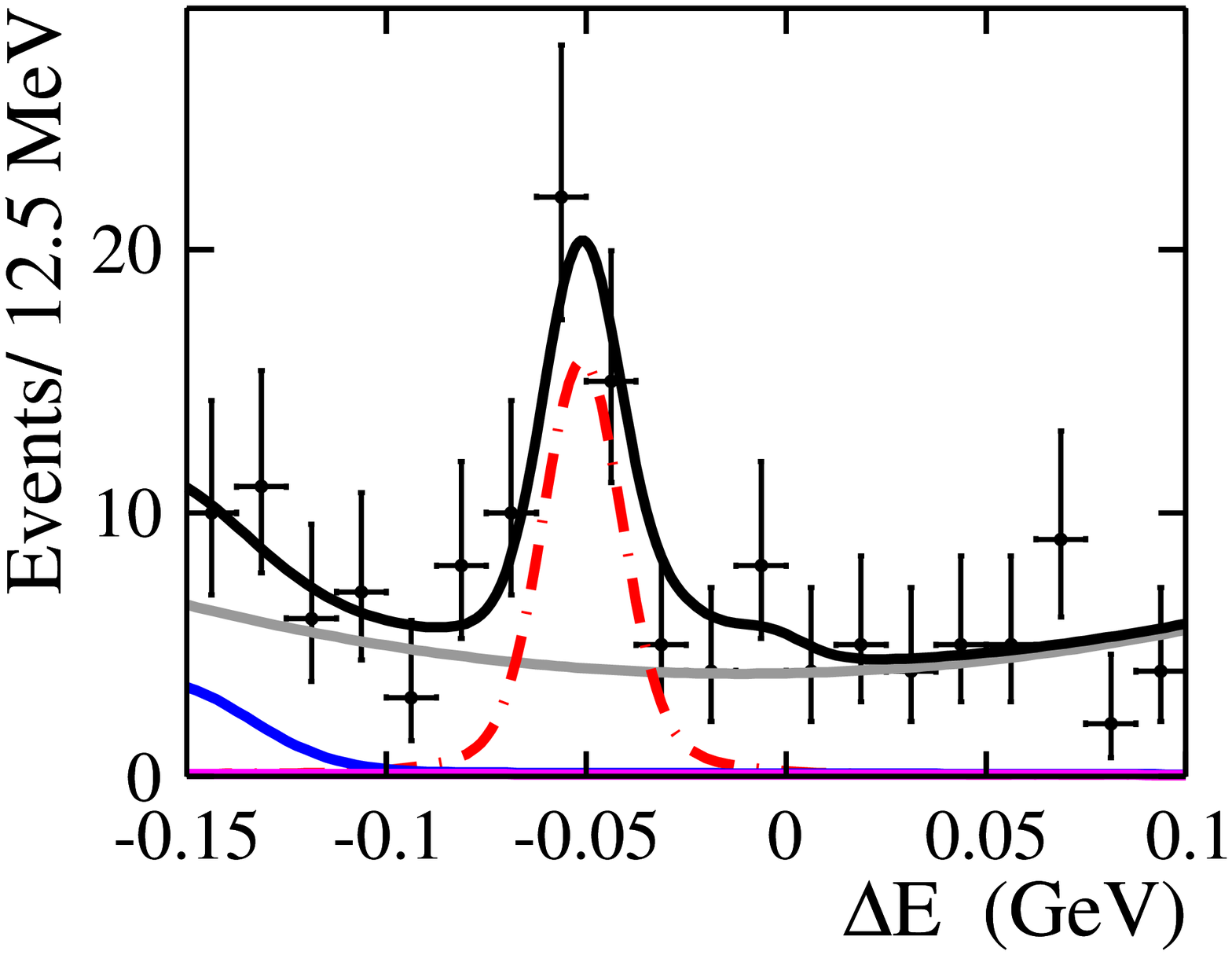,height=1.5in}
\epsfig{file=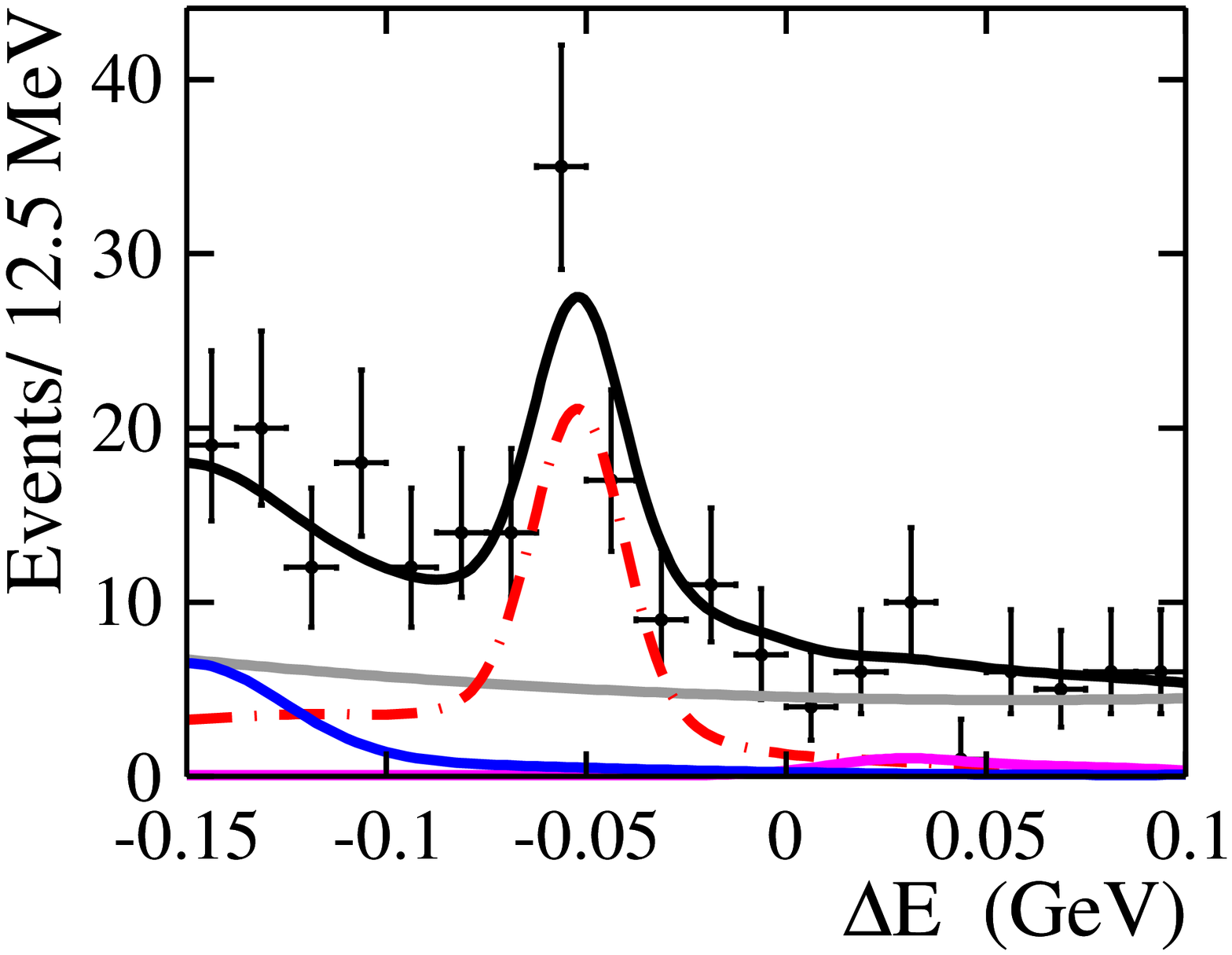,height=1.5in}
\epsfig{file=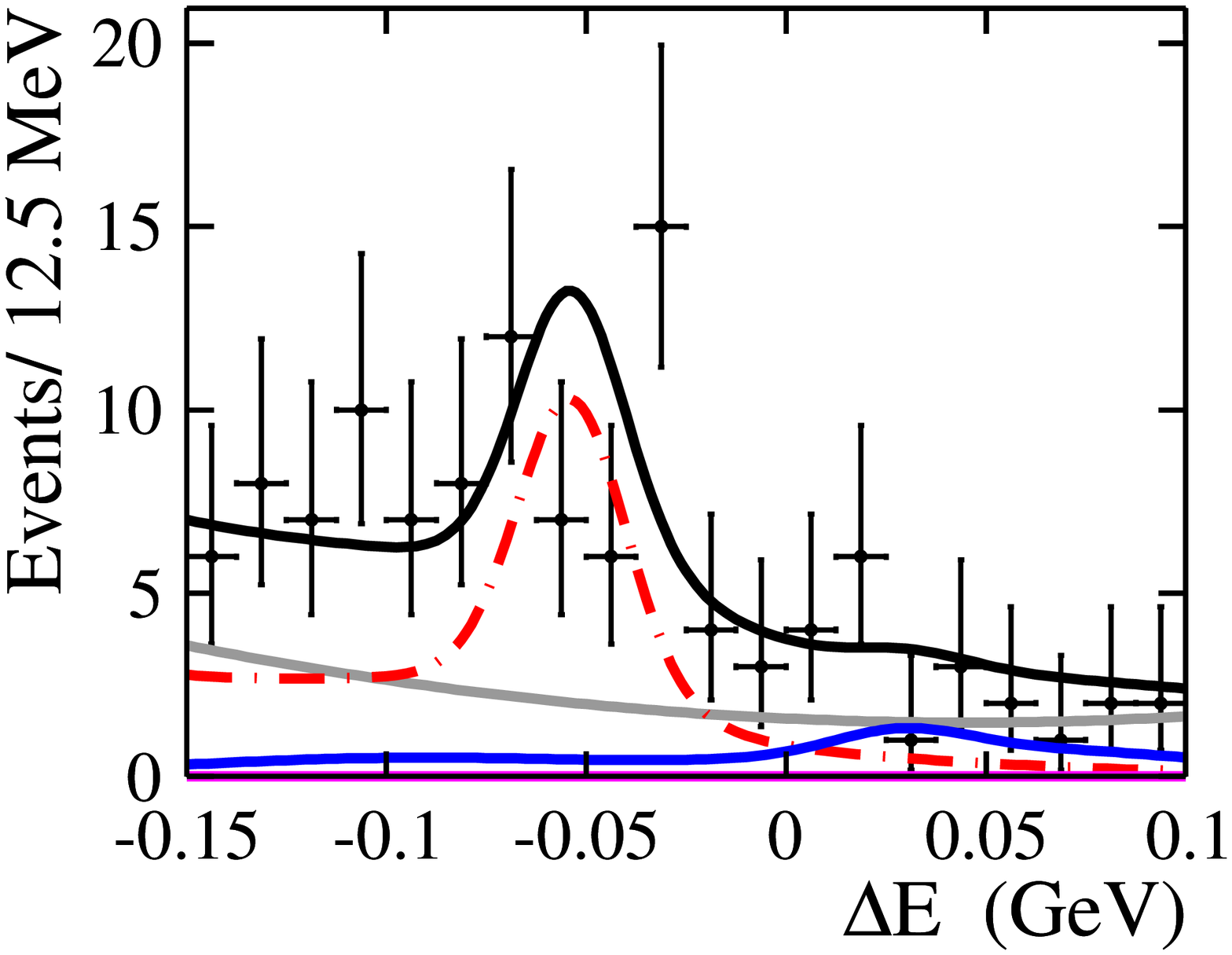,height=1.5in}
\epsfig{file=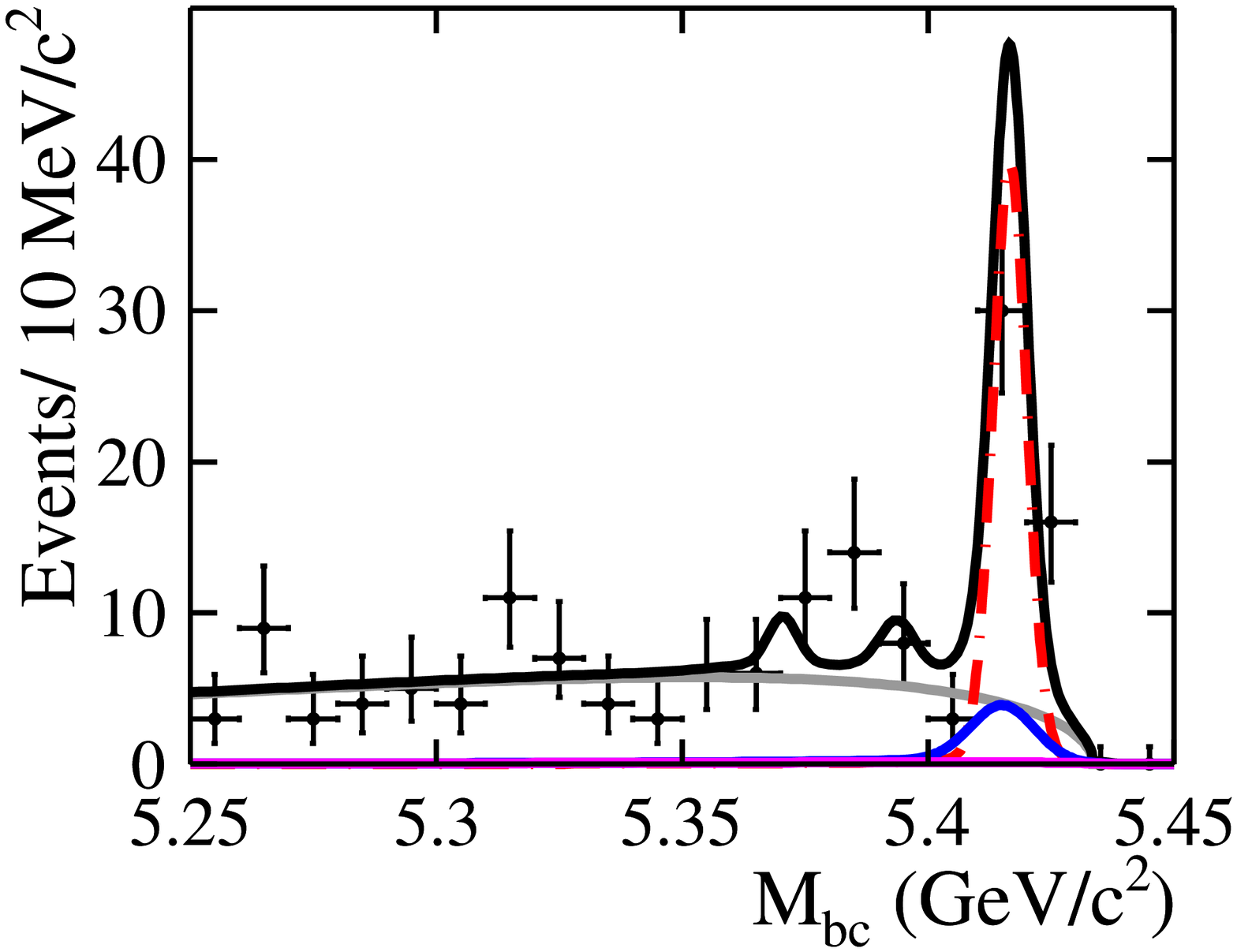,height=1.5in}
\epsfig{file=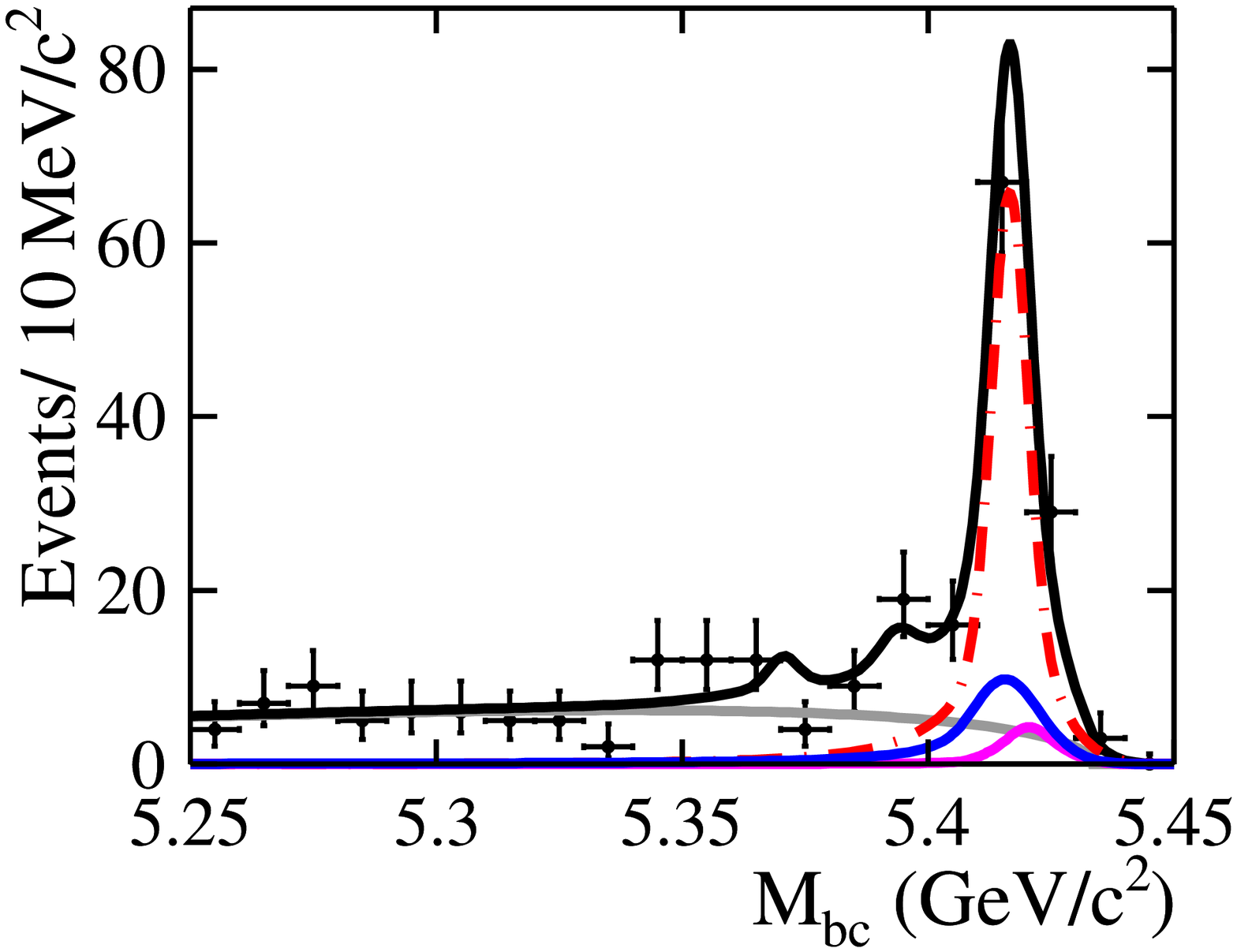,height=1.5in}
\epsfig{file=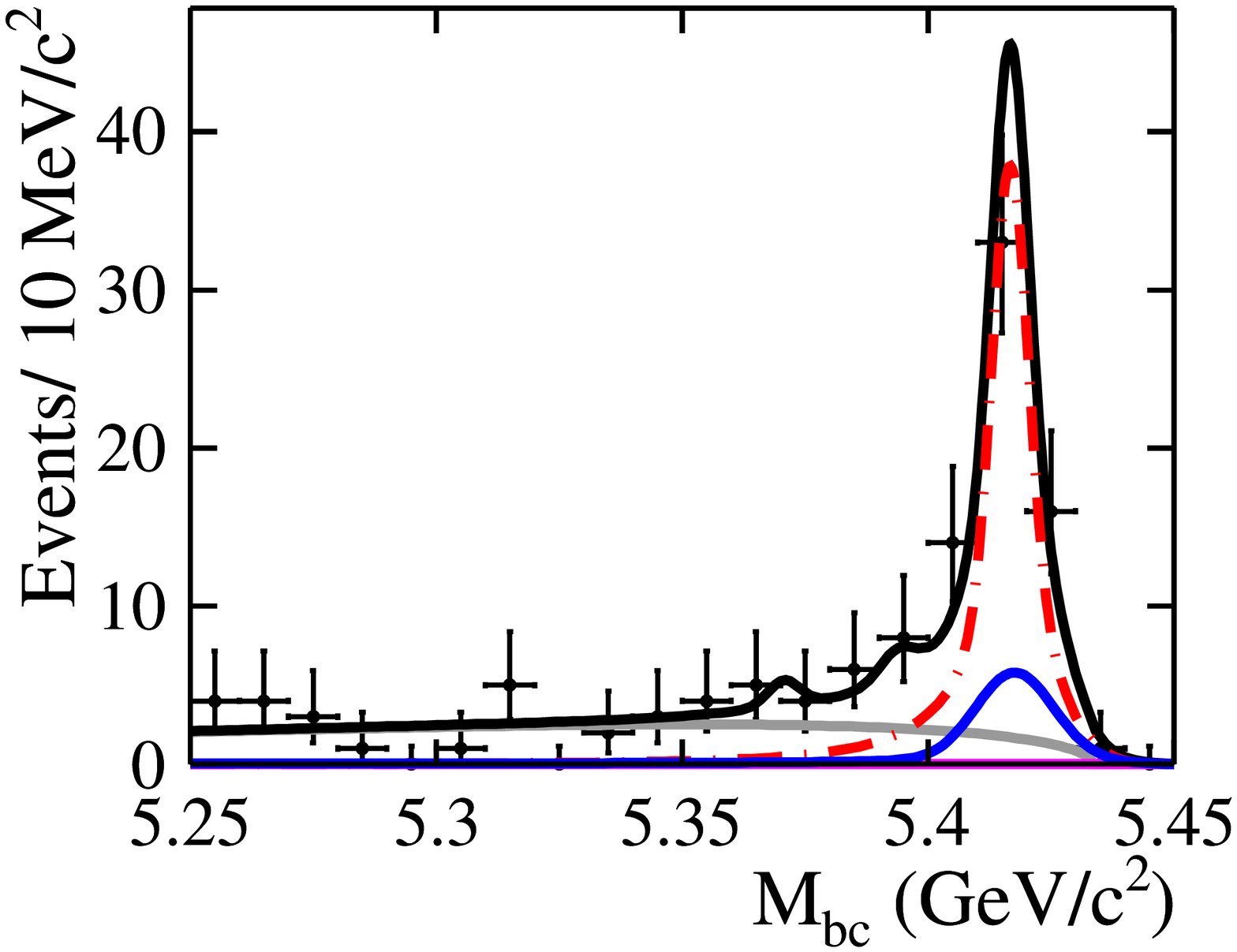,height=1.5in}
\caption{\label{fig:fit_results}
 $\mbc$ and $\de$ projections of the fit result. The columns correspond to
 $B_s^0 \to \dsds$ (left), $B_s^0 \to \dsstds$ (middle), and $B_s^0 \to \dsstdsst$ (right). 
The red dashed curves show CR+WC signal, the blue and purple solid curves
show CF, the grey solid curves show background, and the black solid curves show the total.}
\end{center}
\end{figure}

The projections of the fit are shown in Fig.~\ref{fig:fit_results}
and the fitted signal yields and the branching fractions are listed in Table~\ref{tab:fit_results}. 
The systematic errors are listed in Table~\ref{tab:syst_errors}. 


\begin{table}[bth]
\renewcommand{\arraystretch}{1.2}
\begin{center}
\begin{tabular}{ l | c c c c }
Mode & $Y$(events) & $\varepsilon$ ($\times 10^{-4}$)  & ${\cal B}$ (\%)  & $S$ \\ \hline
$D^+_s D^-_s$ &   $33.1_{-5.4}^{+6.0}$ & 4.72 &  $0.58^{+0.11}_{-0.09}$ $\pm0.13$ &11.5\\  
$D^{*\pm}_s D^{\mp}_s$ &  $44.5_{-5.5}^{+5.8}$ & 2.08 &  $1.8\pm0.2$ $\pm0.4$ &10.1\\  
$D^{*}_s D^{*}_s$ &  $24.4_{-3.8}^{+4.1}$ & 1.01 &  $2.0\pm0.3$ $\pm0.5$ &7.8\\  
Sum &   $102.0_{-8.6}^{+9.3}$ &  &  $4.3\pm0.4$ $\pm1.0$ & \\  \hline
\end{tabular} 
\caption{\label{tab:fit_results} Signal yield ($Y$), efficiency including intermediate 
branching fractions ($\varepsilon$), branching fraction (${\cal B}$), 
and signal significance ($S$) including systematic uncertainty. The 
first errors listed are statistical and the second are systematic.}
\end{center}
\end{table}

\begin{table}[bth]
\renewcommand{\arraystretch}{1.1}
\begin{center}
\begin{tabular}{ l   c c c c  c c  }
&\multicolumn{2}{c}{$D^+_sD^-_s$} & 
\multicolumn{2}{c}{$D^*_s D^{}_s$} & 
\multicolumn{2}{c}{$D^{*+}_s D^{*-}_s$} \\
\hline
 Source &  $+\sigma$ & $-\sigma$ & $+\sigma$ & $-\sigma$ & $+\sigma$ & $-\sigma$ \\
\hline
Signal PDF shape   & 2.7 &2.2   & 2.2& 2.4& 5.1& 3.8 \\
Bckgrnd PDF shape & 1.5 &1.3   & 1.3& 1.4& 2.9& 2.8 \\
WC + CF fraction   & 0.5 & 0.5 & 4.7 & 4.5 & 11.0 & 9.7 \\
$\mathcal{R}$ requirement 
    & 	3.1 & 0.0	& 0.0 & 2.7 & 0.0 & 2.1   \\
Best cand. selection & 5.5 & 0.0	& 1.5 & 0.0 & 1.5 &  0.0   \\
$\pi^\pm/K^{\pm}$ identif.  & 7.0 & 7.0	& 7.0 & 7.0 & 7.0 & 7.0  \\
$K_{S}$  reconstruction      & 1.1   & 1.1 & 1.1 & 1.1 & 1.1& 1.1  \\
 $\pi^{0}$ reconstruction & 1.1 & 1.1 & 1.1 & 1.1 & 1.1 & 1.1 \\
$\gamma$& - &-	& 3.8 & 3.8 & 7.6 & 7.6 \\
Tracking    &  2.2 & 2.2 & 2.2 & 2.2 & 2.2	& 2.2  \\
Polarization   &  0.0 & 0.0 & 0.8 & 2.8 & 0.6	& 0.2  \\
\hline
MC statistics for $\varepsilon$ & 0.2 &  0.2	& 0.4  & 0.4 &  0.5 &  0.5 \\
$\cal{B}(D_{s}^{(*)})$  & 8.6& 8.6& 8.6&8.6& 	8.7& 8.7 \\
$N_{B_s^{(*)}B_s^{(*)}}$            & \multicolumn{6}{c}{18.3} \\
$f^{}_{B^*_s\overline{B}^*_s}$ & \multicolumn{6}{c}{2.0} \\
\hline
Total  & 22.7 & 21.8  &  22.7 & 22.9 & 26.2 & 25.5 \\ \hline
\end{tabular}
\caption{\label{tab:syst_errors}Systematic errors (\%). 
Those listed in the top section affect the signal yield 
and thus the signal significance.}
\end{center}
\end{table}

\section{$\Delta\Gamma_s$ Estimation}

In the heavy quark limit with $(m^{}_b-2m^{}_c)\to 0$ and 
$N^{}_c\to\infty$, the dominant contribution to the decay 
width comes from $\bsdsds$ decays~\cite{Shifman,Aleksan}.
Assuming negligible $CP$ violation, the branching fraction 
is related to $\dgs$ as $\dgs/\gs = 2{\cal B}/(1-{\cal B})$.
Inserting the total ${\cal B}$ from Table~\ref{tab:fit_results} gives
\begin{eqnarray}
\frac{\dgs}{\gs} & = & 
0.090\pm0.009 \,\pm0.023\,,
\label{eqn:dg_result}
\end{eqnarray}
where the first error is statistical and the second is systematic. 
The precision of this result is similar to that of other recent 
measurements~\cite{LHCb-dgs,CDF-dgs} and  consistent with theoretical predictions~\cite{Nierste}. 
Two main uncertainties are the unknown $CP$-odd component in  $B^0\to  D^{*+}_sD^{*-}_s$ decay 
and the size of contributions from three-body final states. 
The former is estimated to be only $6\%$. However
Ref.~\cite{Hou} calculates significant contribution from $B_s^0 \to D_s^{(*)}D^{(*)}K^{(*)}$ decays. 
This calculation predicts $\dgs/\gs$  from $D_s^{(*)+}D_s^{(*)-}$ alone to be $0.102\pm0.030$, 
which agrees well with our result.

\section{Polarization Measurement of $\dsstdsst$}

\begin{figure}[bth]
\begin{center}
\epsfig{file=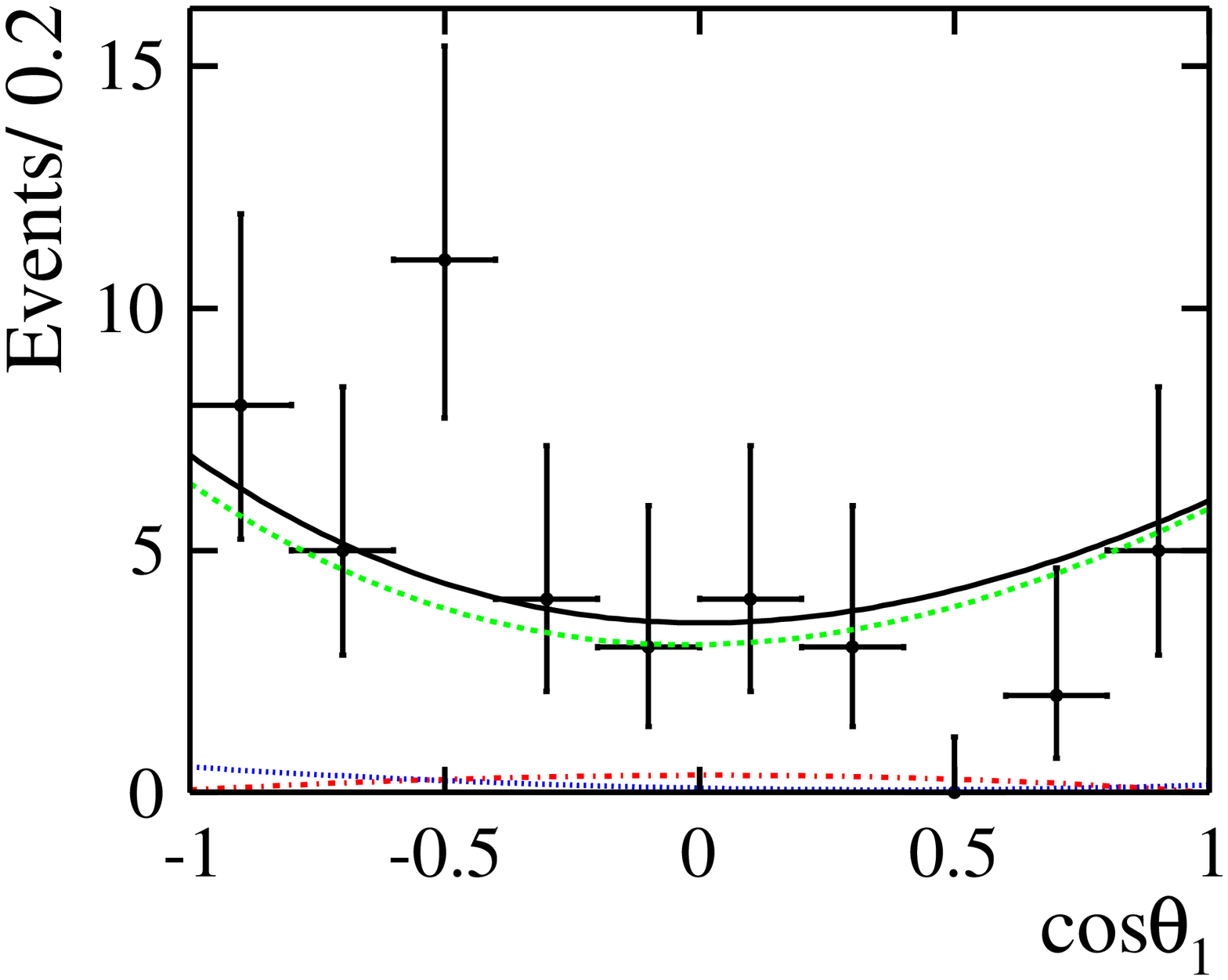,height=1.5in}
\epsfig{file=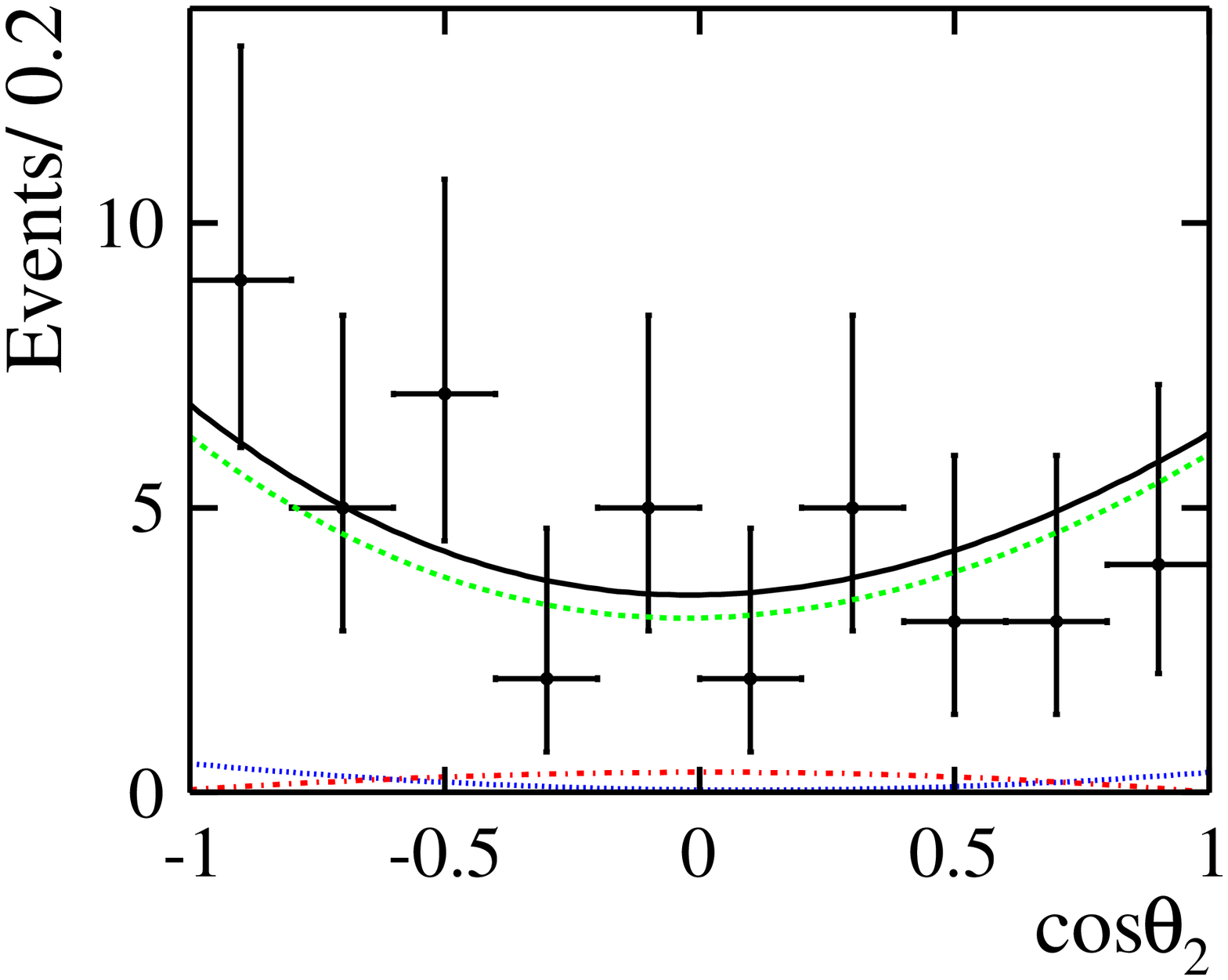,height=1.5in}
\caption{\label{fig:helicity_angles} 
Helicity angle distributions and projections of the
fit result. The green dashed (red dashed-dotted) curves show the 
transverse (longitudinal) components,
the blue dotted curve shows background, and the black solid curve shows the total.}
\end{center}
\end{figure}

We have also made the first measurement of the longitudinal
 polarization fraction ($\fl$)
of $\dsstdsst$.  We select events using the same criteria 
as before, however we use a narrower range of $\mbc$ and $\de$ ($2.5\sigma$ in resolution) in order to  minimize $B_s^0 \to \dsstds$ cross-feed.
For these events we perform an unbinned maximum-likelihood fit to the helicity 
angles $\theta^{}_1$ and $\theta^{}_2$, which are the angles 
between the daughter $\gamma$ momentum and the opposite of the $B_s^0$ momentum in the $D^{*\pm}_s$ rest frame. 
The angular distribution is
$\left(|A_+|^2 + |A_-|^2\right)\left(\cos^2\theta_1 +1\right)
\left(\cos^2\theta_2 +1\right) +
|A^{}_0|^2 4\sin^2\theta^{}_1\sin^2\theta^{}_2$,
where $|A^{}_+|$, $|A^{}_-|$, and $|A^{}_0|$ are the helicity amplitudes.
The fraction $\fl$ is given by $|A^{}_0|^2/(|A^{}_0|^2+|A^{}_+|^2+|A^{}_-|^2)$.
We measure
\begin{eqnarray}
f^{}_L & = & 0.06\,^{+0.18}_{-0.17}\,\pm0.03\,,
\label{eqn:helicity_result}
\end{eqnarray}
where the first error is statistical and the second is systematic. 
The systematic errors are from
signal PDF shapes ($+0.008,-0.010$),
background PDF shape ($+0.007, -0.004$),
fixed WC fractions ($+0.013,-0.015$),
fixed background level ($\pm 0.022$),
continuum suppression ($+0.011$), possible fit bias ($-0.011$), and
MC efficiency due to statistics ($\pm 0.0004$).
The helicity angle distributions and fit projections
are shown in Fig.~\ref{fig:helicity_angles}.

In summary, using the branching fractions of
$\bsdsds$ decays,
 we measured $\dgs/\gs$ assuming no $CP$ violation, 
among other theoretical assumptions. 
We have also made the first
measurement of the $B_s^0 \to \dsstdsst$
longitudinal polarization fraction.



\begin{thebibliography}{99}
\bibitem{charge-conjugates} 
Charge-conjugate modes are implicitly included.
\end{thebibliography}
\end{document}